\documentclass[a4paper,12pt]{article}

\usepackage[dvips]{graphicx}
\usepackage{epsfig}

\newcommand\new{\newcommand}         % shorthand for \newcommand
\new{\as}[1]      {{\ifmmode\alpha^{#1}_s\else$\alpha^{#1}_s$\fi}}
\new{\asmz}       {{\ifmmode\alpha_s(\Mzsq)\else $\alpha_s(\Mzsq)$\fi}}
\new{\Mzsq}       {{\ifmmode\mathrm{ M}^2_{\mathrm{ Z}}\else $\mathrm{ M}^2_{\mathrm{ Z}}$\fi}}
\new{\asmu}[1]    {{\ifmmode\alpha_s^{#1}(\mu^2)\else $\alpha_s^{#1}(\mu^2)$\fi}}
\new{\asmtau}     {{\ifmmode\alpha_s(m^2_{\tau})\else $\alpha_s(m^2_{\tau})$\fi}}
\new{\Oas}[1]     {{\ifmmode{\mathcal O}(\alpha_s^{#1})\else ${\mathcal O}(\alpha_s^{#1})$\fi}}
\new{\LEP}        {\mbox{\textsc{LEP}}}
\new{\CERN}       {\mbox{\textsc{CERN}}}
\new{\ALEPH}      {\mbox{\textsc{ALEPH}}}
\new{\DELPHI}     {\mbox{\textsc{DELPHI}}}
\new{\LD}         {\mbox{\textsc{L3}}}
\new{\OPAL}       {\mbox{\textsc{OPAL}}}

\begin{document}

%%%%%%%%%%%%%%%%%%%%%%%%%%%%%%%%%%%%%%%%%%%%%%%%%%%%%%%%
% The title, all uppercase; if you want to split it in
% two or more lines, put a \\ macro at each line break
% example: 
%   \title{TITLE: FIRST LINE\\ SECOND LINE}
%
\title{EXPERIMENTAL STATUS OF\\ 
       MEASUREMENTS OF $\alpha_s$ AT LEP\footnote{to be
       published in the proceedings of the XIIIth meeting on physics at \LEP,
       LEPTRE, Rome, April 2001.}}

%%%%%%%%%%%%%%%%%%%%%%%%%%%%%%%%%%%%%%%%%%%%%%%%%%%%%%%%
% The author(s), separated by commas; do not put a
% comma before the last author, use instead the \And
% macro which produces a normal ``and'' in the
% caps/small caps context
%
\author{G.\ Dissertori, \\ \small{EP Division, CERN, Geneva}}

%%%%%%%%%%%%%%%%%%%%%%%%%%%%%%%%%%%%%%%%%%%%%%%%%%%%%%%%
%

\maketitle

%%%%%%%%%%%%%%%%%%%%%%%%%%%%%%%%%%%%%%%%%%%%%%%%%%%%%%%%
% Write the text starting from here and using the usual
% LaTeX commands.
% ABSTRACT
\begin{abstract}
  \noindent
  A summary is given of the current status of measurements
  of the strong coupling constant \as{} performed at 
  \LEP. These include measurements from inclusive observables
  as well as from event shape variables. Recent results
  based on power law corrections are discussed.
\end{abstract}

%%%%%%%%%%%%%%%%%%%%%%%%%%%%%%%%%%%%%%%%%%%%%%%%%%%%%%%%
% Introduction
%
\section{Introduction}
Quantum Chromodynamics (QCD) is generally accepted as
the theory of strong interactions between quarks and
gluons. If the quark masses are fixed, there is only
one free parameter in the theory, the strong coupling
constant \as{}. Thus measurements of this parameter are 
of paramount importance. Such measurements have been
performed in a large variety of experiments, for
various initial states, processes and energy scales.
During recent years good consistency has been
obtained between the different determinations
and the theoretical prediction for the running (energy
dependence) of the coupling \cite{PDG, Bethke}. 
Here a summary of
the measurements performed at \LEP\ is given, which 
contribute in an important way to the achieved precision 
on the world average value of $\asmz = 0.118 \pm 0.002$ 
\cite{PDG}.

%%%%%%%%%%%%%%%%%%%%%%%%%%%%%%%%%%%%%%%%%%%%%%%%%%%%%%%%
% Inclusive observables
%
\section{Inclusive observables}

At \LEP\ the strong coupling constant is measured in
$\mathrm{e}^+\mathrm{e}^-$ annihilations into quarks which subsequently
fragment into hadrons, or in 
$\mathrm{e}^+\mathrm{e}^-\rightarrow\tau^+\tau^-$,
where one or both tau leptons decay hadronically.
For inclusive observables such as the total cross 
section the hadronic final state is not analyzed
w.r.t.\ its particular properties such as particle
content or topology. In practice a ratio of cross
sections or partial decay  widths is measured, 
in order to take advantage of the
cancellation of common systematic uncertainties, 
%such as
%\begin{equation}
%  R_{l(\tau)} = \frac{\Gamma\left(Z(\tau)\rightarrow\mathrm{hadrons}\right)}
%                     {\Gamma\left(Z(\tau)\rightarrow\mathrm{leptons}\right)}
% \;\mbox{$=$}\; R_{l(\tau)}^{\mathrm{EW}}\,
% \left(1 + \delta_\mathrm{QCD} + \delta_\mathrm{mass} +
%           \delta_\mathrm{np} \right) \quad .
%\end{equation}
$R_{l(\tau)} = \Gamma\left(Z(\tau)\rightarrow\mathrm{hadrons}\right)/
               \Gamma\left(Z(\tau)\rightarrow\mathrm{muons}\right) = $
$R_{l(\tau)}^{\mathrm{EW}}\,\left(1 + \delta_\mathrm{QCD} + \delta_\mathrm{mass} +
           \delta_\mathrm{np} \right)$.
The QCD correction is known up to
next-to-next-to-leading order (NNLO),
\textit{ie.}, $\delta_\mathrm{QCD} = \sum_{n=1}^{3} c_n \as{n}$,
with known coefficients $c_{1\ldots 3}$. 
In addition there are corrections for quark mass and non-perturbative
effects.

In the case of $R_l$ the \LEP\ combination turns out to 
be $R_l = 20.768\pm 0.024$ \cite{LEPEW}, which
translates into $\asmz = 0.124\pm 0.004$ \cite{Bethke}. 
The error
is dominated by experimental uncertainties, mainly the
statistical error in the muon cross section. Mass corrections
and non-perturbative effects are very small, since they are
suppressed as $m_\mathrm{q}^2/\Mzsq$ and $c/\mathrm{ M}^4_{\mathrm{ Z}}$.
Theoretical uncertainties arise from the missing knowledge
of the Higgs mass and the uncertainty on the top mass,
as well as from renormalization scale and scheme variations. 

For tau decays \ALEPH\ and \OPAL\ have measured the hadronic
mass spectrum and determined its moments. For example,
$R_\tau$ corresponds to the zeroth moment. By measuring
several moments, not only \as{} can be fitted, but at the same
time also some non-perturbative parameters. 
%In fact, the non-perturbative correction
%$\delta_\mathrm{np} = -0.007\pm0.004$ turns out to be small.
A combination of the results from the two experiments
yields
$\asmz = 0.1181\pm 0.0007 (\mathrm{exp}) \pm 0.0030 (\mathrm{theo})$ \cite{Bethke},
which represents one of the most precise measurements. In contrast
to $R_l$, here the experimental error is negligible, whereas
the theoretical uncertainty dominates. This uncertainty stems
from different estimates of higher order corrections.

%A somewhat less inclusive measurement is obtained by 
%studying the scaling violations in fragmentation functions,
%\textit{ie.}, the inclusive momentum spectrum of hadrons
%measured at different centre-of-mass energies. These scaling
%violations are described by Altarelli-Parisi evolution
%equations, known in NLO, which depend on \as{}. 
%A combination of results from \ALEPH\ and \DELPHI\ gives
%$\asmz = 0.126^{+0.006}_{-0.007} (\mathrm{exp}) \pm 0.009 (\mathrm{theo})$ \cite{Bethke} ,
%which is less precise, but nevertheless nicely consistent 
%with other determinations. 

%%%%%%%%%%%%%%%%%%%%%%%%%%%%%%%%%%%%%%%%%%%%%%%%%%%%%%%%
% Event shapes
%
\section{Event shape distributions}

Event shape distributions are observables which are sensitive
to the topology of an hadronic event, and thus give direct
sensitivity to \as{} since gluon radiation influences the
topology. For a particular set of infrared and collinear safe variables
the cross section is known at NLO, and large logarithms
have been resummed to all orders in \as{} (see \cite{Grazzini} for a review on
this topic). 
%Thus it  can be schematically written as \cite{Grazzini}
%\begin{equation}
%  \frac{\mathrm{d}\sigma}{\mathrm{d}y} = \as{} A(y) + \as{2} B(y) + 
%    \sum_{n,m} c_{nm} \as{n} L^M 
%  \quad .
%\end{equation}
This set of observables comprises thrust, heavy jet mass,
the differential two-jet rate computed in the Durham scheme,
the C parameter, as well as the total and wide jet broadening. 
These distributions have been measured by all \LEP\ collaborations and used for
\as{} determinations, for a large
set of centre-of-mass energies at \LEP I and \LEP II (for
a review see \textit{eg.} \cite{Bethke}).
Recently, also the four-jet rate (Durham scheme) has been employed
for \as{} measurements \cite{Bravo, OPAL}, since a NLO calculation
as well as the resummation of large logarithms is at hand. 

For this kind of observables the dominant uncertainties on
\as{} are of theoretical origin. Firstly the purely perturbative
predictions have to be corrected for hadronization effects,
which are larger than for inclusive observables. These corrections
are taken from Monte Carlo programs based on various
phenomenological models, leading therefore to ambiguities.
Secondly, there are uncertainties stemming from the
estimates of missing higher order terms. As an example, in a
recent preliminary analysis by \ALEPH\ \cite{ALEPH01} of the
data at 206 GeV centre-of-mass energy
the following uncertainties on \as{} are quoted when combining
the results from the six event shapes mentioned above~:
$0.0022 (\mathrm{stat})$, $0.0017 (\mathrm{exp})$, 
$0.0007 (\mathrm{had})$, $0.0016 (\mathrm{match})$, 
$0.0038 (\mathrm{scale})$.
``Match'' stands for an ambiguity in the combination of fixed
order and resummation calculations, and ``scale'' for
variations of the renormalization scale. The statistical
error can be substantially reduced by combining with the
results from the Z resonance and from other experiments.
Therefore these findings clearly
indicate that a considerable improvement can only be obtained if the
theoretical uncertainties are reduced. This might
be possible with the advent of NNLO calculations for
three-jet observables.  

A somewhat different approach to the problem of missing
higher orders has been attempted by \DELPHI\ \cite{DELPHI}.
Using only the NLO prediction for a large set of event shape
variables, and setting the renormalization scale equal
to the centre-of-mass energy, a large scatter in the
\as{} values from different observables is found. However,
this scatter is considerably reduced if both \as{} and
the renormalization scale are fitted at the same time.
The price to pay is a large scatter in renormalization
scales, from $5$ to $240$ GeV. These extreme values might not be
natural from a theoretical point of view. In any case,
a clear indication for large missing higher orders
is obtained, which depend strongly on the observable.

%%%%%%%%%%%%%%%%%%%%%%%%%%%%%%%%%%%%%%%%%%%%%%%%%%%%%%%%
% Power laws
%
\section{Power law corrections}

During recent years there has been a lot of activity 
on the phenomenology of power law corrections to event
shape variables. For an overview the interested reader
is referred to \cite{Banfi}. Instead of obtaining the
non-perturbative corrections from Monte Carlo models 
it turns out that additive terms in the case of moments
or shifts in the case of event shape distributions of the
type $1/Q$ are capable of giving a good description of
the data. Using this power law behaviour, \as{} can
be determined together with a non-perturbative parameter
$\alpha_0$ which is expected to be universal. A recent
analysis of a large set of data from a wide range
of energies \cite{Kluth} indicates that indeed universality
is obtained at the 20\% level for the mean values of
event shapes. However, in the case of distributions 
still some larger fluctuations are observed. Several
effects might be at the origin of this. It might be
that hemisphere variables such as heavy jet mass and
jet broadenings need a more careful analysis than
event variables such as thrust. An approach in this
direction could be the introduction of a non-perturbative
shape function (see \cite{Tafat} for a recent review).
In addition, further understanding is needed for the
effects of hadron masses and resonance decays \cite{Salam}.
In any case, a high precision measurement of \as{}
based on power law corrections will need further
theoretical as well as experimental investigations.

\begin{figure}[thb]
 \begin{center}
 \begin{tabular}{cc}
   \includegraphics[width=7.5cm]{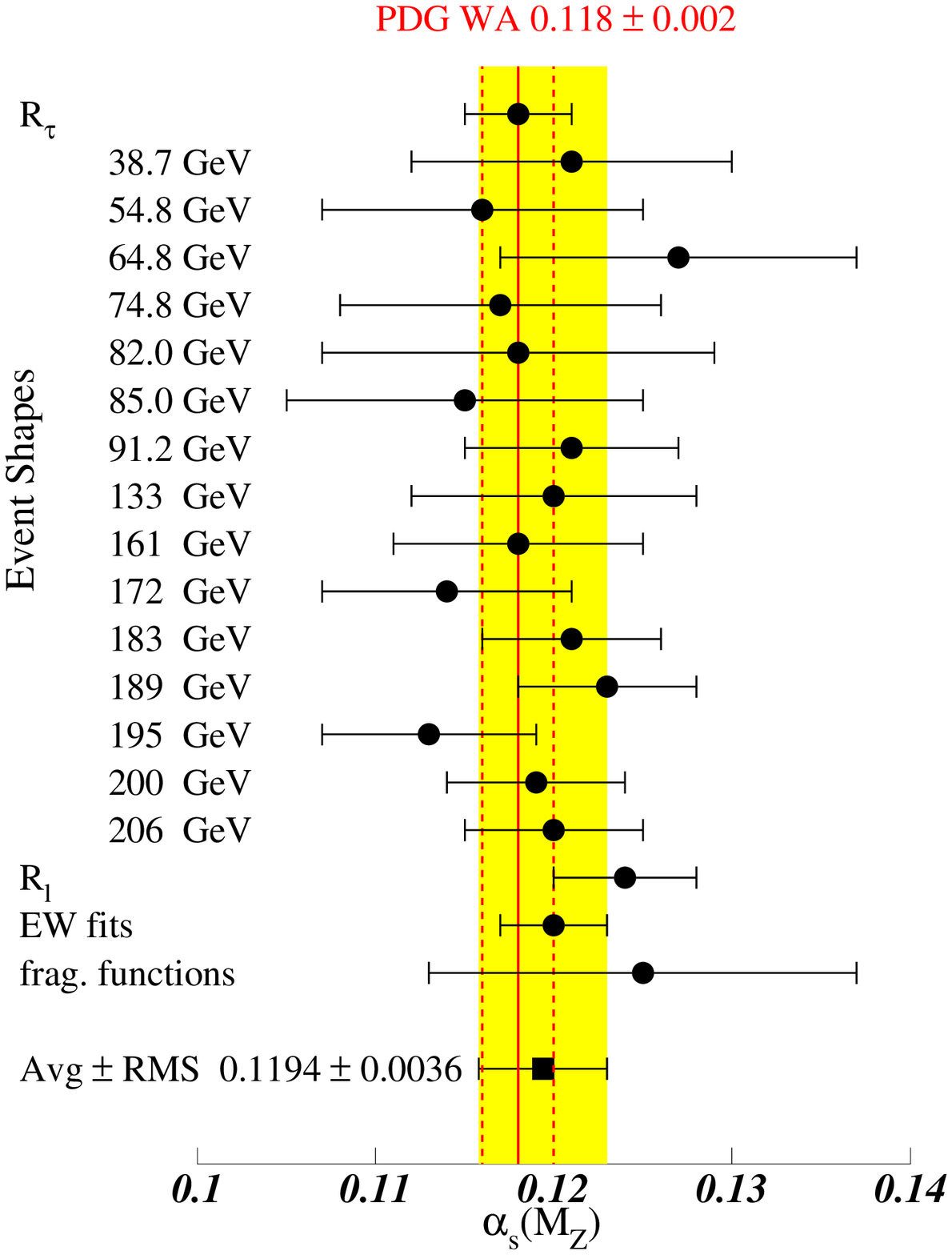} &
   \includegraphics[width=7.5cm]{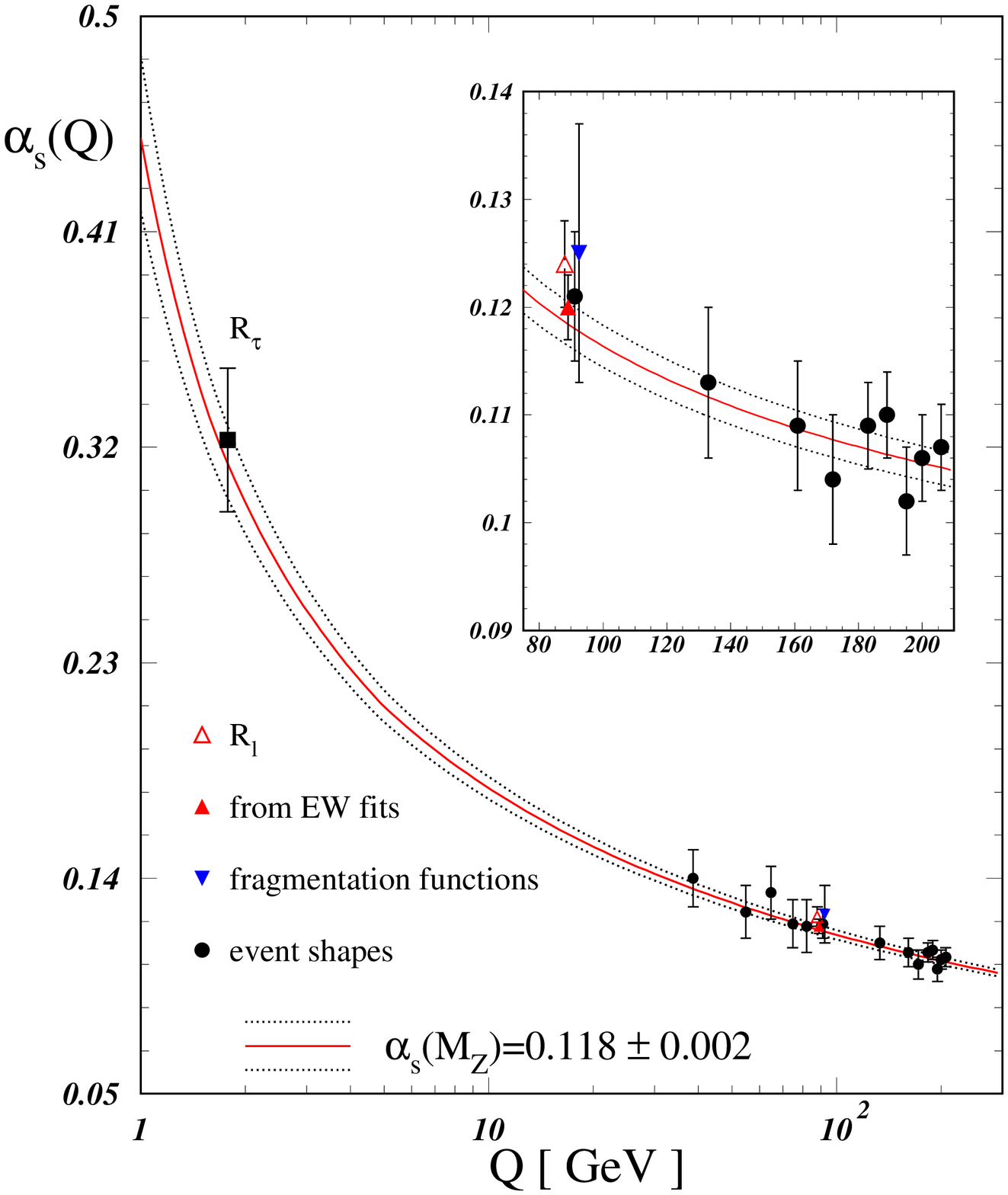} 
 \end{tabular}
 \caption{Summary of the measurements of \as{} performed
  at \LEP. \label{summary}}
 \end{center}
\end{figure}

%%%%%%%%%%%%%%%%%%%%%%%%%%%%%%%%%%%%%%%%%%%%%%%%%%%%%%%%
% Summary
%
\section{Summary}

A large set of measurements of \as{} is available from
\LEP. These determinations contribute in an important
way to the world average value and are nicely consistent
with the predicted running of the strong coupling, as
shown in fig.\ \ref{summary}. The results from event shape
variables have shown that further theoretical work
is needed in order to improve the precision, such as 
the completion of NNLO calculations for three-jet quantities
and a better understanding of power law corrections. 
Since these developments  might not be finished within
a few years from now, it is of great importance
to preserve the \LEP\ data well, in order to facilitate a 
possible re-analysis in the future.

%%%%%%%%%%%%%%%%%%%%%%%%%%%%%%%%%%%%%%%%%%%%%%%%%%%%%%%%
% Thanx
%
\section{Acknowledgements}

I would like to thank S.~Kluth and S.~Tafat for 
providing me with useful information on power law
corrections.

%%%%%%%%%%%%%%%%%%%%%%%%%%%%%%%%%%%%%%%%%%%%%%%%%%%%%%%%
% bibliography
%

%%%%%%%%%%%%%%%%%%%%%%%%%%%%%%%%%%%%%%%%%%%%%%%%%%%%%%%%
% End of the paper
%
\end{document}